\documentclass[draft,preprint,amsmath,showpacs]{revtex4}
\begin{document}

\title{Enhancement of Persistent Current in Metal Rings by Correlated Disorder }
\author{Jean Heinrichs}
\email{J.Heinrichs@ulg.ac.be} \affiliation{D\'{e}partement de
Physique, B{5a}, Universit\'{e} de Li\`{e}ge, Sart Tilman,
B-{4000} Li\`{e}ge, Belgium}
\date{\today}
\begin{abstract}
We study analytically the effect of a correlated random potential on
the persistent current in a one-dimensional ring threaded by a
magnetic flux $\phi$, using an Anderson tight-binding model.  In our
model, the system of $N=2M$ atomic sites of the ring is assumed to
be partitioned into $M$ pairs of nearest-neighbour sites (dimers).
While the individual atomic site energies are assumed to be
identically distributed gaussian variables with autocorrelation
parameter $\varepsilon^2_0$, the dimer site energies are chosen to
be correlated with a gaussian strength $\alpha^2<\varepsilon^2_0$.
For this system we obtain the exact flux-dependent energy levels to
second order in the random site energies, using an earlier exact
transfer matrix perturbation theory.  These results are used to
study the mean persistent current generated by $N_e\leq N$ spinless
electrons occupying the $N_e$ lowest levels of the flux-dependent
energy band at zero temperature. Detailed analyzes are carried out
in the case of low filling of the energy band ($1\ll N_e\ll N$) and
for a half-filled band ($N_e=N/2$), for magnetic fluxes $-1/2
<\phi/\phi_0<1/2$. In the half-filled band case, the uncorrelated
part of the disorder reduces the persistent current while the
correlated part enhances it, in such a way that for
$\alpha^2<\varepsilon^2_0/2$ the current decreases with the
disorder, while for $\alpha^2>\varepsilon_0^2/2$ it increases with
it. Also, while showing a specific dependence on the flux, the
disorder effect has the same dependence on the parity of $N_e$ as
the pure system free electron current.  In contrast, at low filling
of the energy-band, the disorder induced effect in the persistent
current depends critically on the parity: due to a peculiar
dependence on the flux, it yields a reduction of the current for odd
$N_e$ and an enhancement of it for even $N_e$. The observability of
the effects of weak correlated disorder on persistent current in the
half-filled band case is restricted to ring sizes in the nanoscale
range, for which no measurements presently exist.
\end{abstract}

\pacs{72.15.Rn,73.23-b,73.23.Ra,73.63.Nm}

\maketitle

\section{INTRODUCTION}

The study of persistent current in small metallic rings threaded by
a magnetic flux has a long history, starting with the basic papers
of London\cite{23} and Hund\cite{24}.  It has been boosted in recent
years with the appearance of the seminal paper of B\"{u}ttiker, Imry
and Landauer\cite{1} for one-dimensional rings. Important aspects
related to the effects of disorder and of electron-electron
interaction on the persistent current have been reviewed and
compared with experimental observations\cite{2,3,4} in well-known
monographs\cite{5,6}and in review articles\cite{7}. An important
unsolved problem remains however to reconcile theoretical and
experimental results for the persistent current, particularly in the
case where the current averaged over many experimental realizations
of isolated disordered rings of a given metal has been
measured\cite{2}. In this case the theoretical results for the
ensemble-averaged current are between one or two- orders of
magnitude lower\cite{5} than the experimental values\cite{2}
depending on whether electron-electron interaction effects are
included or not. However, the electron interaction effects are not
easy to calculate and no definitive answer as to their precise role
seems to exist since there are even models in which Coulomb
interactions act to reduce the persistent current rather than to
enhancing it\cite{8}.

In the present paper we examine another effect which may
substantially enhance the persistent current in a disordered ring
subjected to a magnetic flux.  This effect exists when the
disordered potential is not perfectly random as a result of
short-range correlations.  More precisely, we consider an Anderson
tight-binding model of a ring with an even number $N=2M$ of one
level atomic sites (of spacing $a=1$) whose energies $\varepsilon_n$
fluctuate randomly about a fixed free atom level chosen as zero of
energy.  As usual the $N$ site energies are assumed to be
identically distributed independent random variables.  In order to
account for short-range correlations we extend the model as follows:
we divide the ring up into pairs of nearest-neighbour atomic sites
(dimers), namely the pairs (1,2), (3,4) ... (2M-1, 2M), where the
sites within individual pairs are assumed to be gaussian correlated,
whereas sites belonging to different pairs are uncorrelated.
Including the usual autocorrelation, the Gaussian correlation of
site energies is then described by the averages

\begin{equation}\label{eq1}
\langle\varepsilon_n \rangle=0,\;
\langle\varepsilon_m\varepsilon_n\rangle =
\varepsilon^2_0\delta_{\text{min}}+\alpha^2
\left(\delta_{m,2p-1}\delta_{n,2p}+\delta_{m,2p}\delta_{n,2p-1}\right)\quad.
\end{equation}
\noindent  The intersite coupling term in \eqref{eq1} describes
short-range structural (dimer) correlations of an otherwise
uncorrelated (white-noise) random potential.  In the following we
refer to the Gaussian model \eqref{eq1} as an Anderson random dimer
model (ARDM) which was first discussed in the context of
localization on a linear chain\cite{9}, in the case
$\alpha^2=\varepsilon^2_0$. It differs from the random dimer model
(RDM) introduced earlier by Phillips and co-workers\cite{10,11} for
demonstrating the existence of delocalized states in a
one-dimensional system with correlated disorder.  This model
considers chains of $N$ lattice sites composed of clusters of host
atoms $a$ of site energy $\varepsilon_a$ separated by clusters of
defect atoms of energy $\varepsilon_b$ occurring randomly but in
pairs of nearest neighbours\cite{10,11}. Apart from its binary
character this system differs from our ARDM in that clusters of
$a$-atoms are not restricted to even numbers of atoms unlike the
clusters composed of $b$-atoms.  The RDM model of Phillips {\it et
al.} leads to the existence of $\sqrt N$ delocalized states in the
linear chain lattice\cite{10,11}.  In contrast, the ARDM model (for
$\alpha^2=\varepsilon^2_0$) on a linear chain lattice leads to
localized states only, with a localization length $\xi$ given
by\cite{9}

\begin{equation}\label{eq2}
\frac{1}{\xi}\simeq\frac{\varepsilon^2_0 E^2}{4(4-E^2)}
=\frac{E^2}{2\xi_0}\quad,
\end{equation}
for weak disorder.  Here $E=2\cos k$ is the energy band (in units of
a constant nearest-neighbour hopping parameter $V$) of states of
wavenumber $k$ of the ordered lattice and $\xi_0$ is the weak
disorder Thouless localization length for uncorrelated Gaussian
disorder in one dimension.  The expression (\ref{eq2}) is valid at
energies $E$ except for some special values within the energy band
(including $E=0$) where the weak disorder treatment must be
corrected for Kappus-Wegner type anomalies\cite{9,12}.  It follows
from (\ref{eq2}) that the localization length near the band centre
is strongly increased with respect to $\xi_0$ by the effect of
disorder correlation.

For the tight-binding system on a ring threaded by a magnetic flux
in the direction perpendicular to the ring the Schr\"{o}dinger
equation reduces to the set of difference equations\cite{13,14}

\begin{equation}\label{eq3}
-e^{i\frac{2\pi}{N}\frac{\phi}{\phi_0}}\varphi_{n+1}
-e^{-i\frac{2\pi}{N}\frac{\phi}{\phi_0}}\varphi_{n-1}
+\varepsilon_n\varphi_n =E\varphi_n, \quad n=2,3,\ldots ,
N-1\quad,
\end{equation}

\begin{equation}\label{eq4}
-e^{i\frac{2\pi}{N}\frac{\phi}{\phi_0}}\varphi_2
-e^{-i\frac{2\pi}{N}\frac{\phi}{\phi_0}}\varphi_N
+\varepsilon_1\varphi_1 =E\varphi_1\quad,
\end{equation}

\begin{equation}\label{eq5}
-e^{i\frac{2\pi}{N}\frac{\phi}{\phi_0}}\varphi_1
-e^{-i\frac{2\pi}{N}\frac{\phi}{\phi_0}}\varphi_{N-1}
+\varepsilon_N\varphi_N =E\varphi_N\quad,
\end{equation}
which embody the familiar flux-modified (twisted) boundary
condition\cite{1,14} describing the effect of the magnetic flux.
Here $\phi_0=hc/e$ is the flux quantum (with $h$ the Planck
constant, $c$ the velocity of light and $-e$ the electron charge),
$\varphi_n$ is the amplitude of an eigenstate wavefunction at site
$n$, $E$ and $\varepsilon_n$, are the corresponding eigenvalue and
site energies in units of minus a constant nearest-neighbour
hopping parameter.  The equilibrium persistent current carried by
the $k$-th one-electron eigenstate of (\ref{eq3}-\ref{eq5}) is

\begin{equation}\label{eq6}
I_k=-c\frac{\partial E_k}{\partial\phi}\quad,
\end{equation}
and the total persistent current in a ring containing $0< N_e\leq
N$ spinless electrons in the zero temperature ground state is
given by

\begin{equation}\label{eq7}
I= \sum_k I_k\quad,
\end{equation}
where the summation extends over the $N_e$ lowest one-electron
eigenstates of (\ref{eq3}-\ref{eq5}).

Persistent current has been studied recently in the RDM of Dunlap
{\it et al.}\cite{10} in several papers\cite{15,16,17,18}.  In
particular, Liu {\it et al.}\cite{15} have shown that in the model
of Dunlap {\it et al.} the persistent current is enhanced by the
disorder correlation to values approaching the current in an
ordered system if the Fermi level coincides with the energy of a
delocalized state.  On the other hand, their numerical
calculations indicate that the persistent current is strongly
suppressed when no delocalized state is present near the Fermi
level\cite{15}.

The study of the persistent current in the ARDM is of interest in
view of its specific localization properties discussed above.   In
particular this study may allow us to clarify whether the presence
of a correlation induced delocalized state at the Fermi-level is
indispensable for obtaining large persistent currents as discussed
by the authors of Refs \cite{15} and \cite{16} for the RDM.  These
are our main motivations for the present work.  The
ensemble-averaged energy eigenvalues for the ARDM model for weak
disorder are discussed in Sect. \ref{the perturbed energy levels of
the ring}, using the general perturbation expressions for the
eigenvalues of Eqs. (3-5) derived in Ref. \cite{14}.  The detailed
study of the ensemble-averaged persistent current successively for a
system of $N_e\ll N$ spinless electrons occupying a small portion of
the energy-band of the ring described by (\ref{eq3}-\ref{eq5}), and
for a system of $N_e=N/2$ electrons occupying the lower half of the
energy-band at zero temperature, is presented and discussed in Sect.
\ref{the persistent current in the ring}.  Some final remarks are
given in Sect. \ref{concluding remarks}.

\section{THE PERTURBED ENERGY LEVELS OF THE RING}\label{the perturbed energy levels of the ring}

Exact general expressions for the one-electron energy levels of
the ring threaded by a magnetic flux and perturbed by a weak site
energy disorder have been derived from (3-5) in Ref. \cite{14},
using an exact transfer matrix formalism and expanding the
transfer matrix to second order in the disorder.  In the absence
of disorder the energy levels are given by the familiar
tight-binding results\cite{13,14},

\begin{equation}\label{eq8}
E^0_k (\phi)=E^0_k=-2\cos \frac{2\pi}{N} \Biggl
(k+\frac{\phi}{\phi_0}\Biggr ) \quad,
\end{equation}
where $k=0, \pm 1, \pm 2,\ldots$, and the solution of the
eigenvalue equation yields the following exact expressions for the
first- and second order perturbations of the energy
levels\cite{14}

\begin{equation}\label{eq9}
E_k= E^0_k+ E^{(1)}_k+ E^{(2)}_k+\ldots
\end{equation}

\begin{equation}\label{eq10}
E^{(1)}_k\equiv E^{(1)}=\frac{1}{N}\sum^N_{n=1}\varepsilon_n\quad,
\end{equation}

\begin{equation}\label{eq11}
\begin{split}
E^{(2)}_k= \frac{1}{N\sin q_k \sin N q_k} \sum^N_{n=2}
\sum^{n-1}_{m=1}
(E^{(1)}-\varepsilon_m)(E^{(1)}-\varepsilon_n)\\
\times\sin (n-m)q_k\dot\sin (N-n+m)q_k\quad,
\end{split}
\end{equation}
where
\begin{equation}\label{eq12}
q_k=\frac{2\pi}{N} \Biggl ( k+\frac{\phi}{\phi_0}\Biggr ) \quad.
\end{equation}
\noindent As recalled in Ref. \cite{14} the weak disorder expansion
breaks down at flux values equal to half-integer multiples of
$\phi_0$ including $\phi=0$ where (\ref{eq11}) diverges\cite{22}.
Such divergencies are familiar in other perturbative studies of
persistent current\cite{25} and are due to the degeneracies of the
unperturbed electron energies \eqref{eq8} at these flux values.

As in our earlier work\cite{14} we are interested in the change of
persistent current induced by the energy correction (\ref{eq11})
averaged over the disorder.  The derivation of the closed form of
(\ref{eq11}) averaged over the correlated disorder defined by
(\ref{eq1}) involves the following steps:

\begin{enumerate}
\item for each one of the four disorder factors $(E^{(1)})^2,
-\varepsilon_m E^{(1)}, E^{(1)}\varepsilon_n$ and $\varepsilon_m
\varepsilon_n$ (with $E^{(1)}$ defined by (\ref{eq10})) which
enter in the site summations in (\ref{eq11}), we identify
explicitly the individual nearest-neighbour pair terms with a
non-vanishing correlation (\ref{eq1}), in addition to the
autocorrelation terms.

\item after performing the disorder averaging, using (\ref{eq1}),
the various contributions reduce to combinations of geometric
series over sites, which are then summed in closed form. As a
guide to our calculations we list in the appendix the final
explicit results obtained for the four distinct disorder terms in
(\ref{eq11}).  The sum of the contributions (A2-A5) leads, after
further reduction to the following exact expression for the
averaged disorder effect in the ARDM:

\begin{equation}\label{eq13}
\langle E^{(2)}_k\rangle= \langle
E^{(2)}_k\rangle^{\text{uncorr}}+ \langle
E^{(2)}_k\rangle^{\text{corr}}\quad,
\end{equation}
where
\begin{equation}\label{eq14}
\langle E^{(2)}_k\rangle^{\text{uncorr}}=
\frac{\varepsilon^2_0}{4\sin q_k} \Biggl [ \Biggl ( 1+\frac{1}{N}
\Biggr ) \cot \frac{2\pi\phi}{\phi_0} -\frac{1}{N} \cot q_k \Biggr
] \quad,
\end{equation}
and
\begin{multline}\label{eq15}
\langle E^{(2)}_k\rangle^{\text{corr}}= \frac{\alpha^2}{4} \Biggl \{
2 \Biggl ( \cos q_k -\cot\frac{2\pi \phi}{\phi_0} \sin q_k \Biggr )
-\frac{3}{N} \frac{\cot (2\pi\phi/\phi_0)}{\sin q_k}
\Biggr.\\
-\frac{1}{N^2 \sin\frac{2\pi\phi}{\phi_0}\sin q_k} \Biggl [
\frac{e^{-i\frac{2\pi\phi}{\phi_0}}}{1-e^{-2i q_k}} \Biggl (
\frac{i\left (1-e^{i\frac{4\pi\phi}{\phi_0}}\right)}{\sin 2 q_k}
\Biggr .
\Biggr .\\
+\frac{1}{1-e^{2i q_k}} \Bigl ( 1+e^{2 i q_k} -e^{\frac{i
4\pi\phi}{\phi_0}} \Bigr) -\frac{1}{1-e^{4i q_k}}
e^{i\frac{4\pi\phi}{\phi_0}}
(1+e^{6 iq_k})\\
\Biggl. \Biggl . \Biggl . -3\frac{N}{2} -1-(N-1) e^{4 i q_k}
-\frac{N}{2} e^{-4 i q_k} \Biggr ) +\text{c.c.} \Biggr ]
\Biggr\}\quad.
\end{multline}

Here $\langle E^{(2)}_k \rangle^{\text{uncorr}}$ represents the
averaged energy correction resulting from the random potential in
the absence of correlation, which has been studied
earlier\cite{14}, and $\langle E^{(2)}_k \rangle^{\text{corr}}$
represents the effect of the correlation of the disorder described
by the dimer model of Ref. \cite{9}.

The equations (\ref{eq9}) and (\ref{eq13}-\ref{eq15}) are used in
the following section for studying the ensemble-averaged
persistent currents.

\end{enumerate}

\section{THE PERSISTENT CURRENT IN THE RING}\label{the persistent current in the ring}

We first recall the results for the persistent current in the pure
(non-disordered) tight-binding ring obtained from Eqs
(\ref{eq6}-\ref{eq8}).  The energy levels (\ref{eq8}) occupied by
the $N_e$ spinless electrons differ for odd and for even $N_e$, as
well as for different domains of relative magnetic flux
$\phi/\phi_0$, as is well known\cite{13}.  The sets of occupied
$k$-levels entering in the calculation of the total current in
various cases may be inferred from the set of intersecting
parabolas representing the energy level spectrum as a function of
flux in the free electron limit (see e.g. fig. 2 in Ref.
\cite{13}).  They are defined by the values $k=0, \pm 1,\pm
2,\ldots,\pm (N_e-1)/2$ for odd $N_e$ and $-1/2<\phi/\phi_0<1/2$;
and by $k=0,\pm 1,\ldots,\pm (N_e/2-1),-N_e/2$ for
$0<\phi/\phi_0<1/2$ and by $k=0,\pm 1,\ldots,\pm (N_e/2-1), N_e/2$
for $-1/2<\phi/\phi_0< 0$ for even $N_e$, respectively.  The
summation of the contributions of the various levels (\ref{eq8})
reduces to geometric progressions leading to the following mostly
by well-known results for pure system currents
$I^{(0)}$\cite{13,15}:

\begin{equation}\label{eq16}
\begin{split}
I^{(0)}=-I_0 \frac{\sin (2\pi\phi/N\phi_0)}{\sin\pi/N} \simeq -I_0
.\frac{2\phi}{\phi_0}\quad ,\\
\left( \text{odd } N_e,\quad -\frac{1}{2}<\phi/\phi_0 <\frac{1}{2}
\right) \quad ,
\end{split}
\end{equation}

\begin{subequations}\label{eq17}
\begin{align}
I^{(0)} &=-I_0 \frac{\sin (\pi/N)(2\phi/\phi_0-1)}{\sin \pi/N}
\simeq -I_0 \Biggl (\frac{2\phi}{\phi_0}-1
\Biggr )\quad ,\nonumber\\
& \left ( \text{even } N_e,\quad 0<\phi/\phi_0<\frac{1}{2}
\right )\quad ,\label{eq17a}\\
I^{(0)} &=-I_0 \frac{\sin (\pi/N)(2\phi/\phi_0+1)}{\sin \pi/N}
 \simeq -I_0
\Biggl (\frac{2\phi}{\phi_0}+1
\Biggr )\quad ,\nonumber\\
& \left ( \text{even } N_e,-\frac{1}{2}<\phi/\phi_0<0 \right
)\quad ,\label{eq17b}
\end{align}
\end{subequations}

\begin{equation}\label{eq18}
I_0= \frac{ev_F}{N},\quad v_F=\frac{4\pi}{h}\sin k_F,\quad
k_F=\frac{\pi N_e}{N}\quad,
\end{equation}
where the approximate limiting forms correspond to the limit
$N\rightarrow\infty$ and coincide in the case of (\ref{eq16}) and
(\ref{eq17a}) with earlier analytical results for free
electrons\cite{13}. $v_F$ and $k_F$ in (\ref{eq18}) denote the
tight-binding Fermi velocity and the Fermi momentum,
respectively\cite{14}.  The limiting free electron expressions in
(\ref{eq16}) and (\ref{eq17a},\ref{eq17b}) exhibit the familiar
sawtoothed currents shown in figs. (3.a) and (3.b) of Ref.
\cite{13} (whose legends must be interchanged!).

We now study the effect of the correlated disorder on the
persistent current obtained from (\ref{eq6}-\ref{eq10}) and
(\ref{eq13}-\ref{eq15}) for a system of $N_e$ spinless electrons
in the $T=0$ ground state.  Our general discussion is made
systematic by considering successively the case of low filling,
$N_e\ll N$, of the energy band and the case of higher filling
exemplified by the half-filled band case, $N_e=N/2$.

\subsection{LOW FILLING OF ENERGY BAND}\label{low filling of energy band}

For occupied levels with quantum numbers $k\ll N$ and magnetic
flux restricted by (\ref{eq16}) and (\ref{eq17a},\ref{eq17b}) we
use

\begin{equation}\label{eq19}
\sin q_k\simeq
q_k=\frac{2\pi}{N}\left(k+\frac{\phi}{\phi_0}\right)\quad ,
\end{equation}
and by expanding (\ref{eq14}-\ref{eq15}) to lowest order for large
$N$ we obtain

\begin{equation}\label{eq20}
\langle E^{(2)}_k\rangle= \frac{\varepsilon^2_0}{4 q_k} \Biggl (
\cot\frac{2\pi\phi}{\phi_0}-\frac{1}{N q_k} \Biggr )
-\frac{3\alpha^2}{4N q^2_k}+ O [(q_k)^0=1]\quad ,
\end{equation}
where

\begin{equation}\label{eq21}
\langle E^{(2)}_k\rangle^{\text{uncorr}}=
\frac{\varepsilon^2_0}{4q_k} \Biggl (
\cot\frac{2\pi\phi}{\phi_0}-\frac{1}{N q_k} \Biggr ) \quad ,
\end{equation}
is the uncorrelated disorder perturbation at low band-filling
obtained from (\ref{eq14}).  The Eq. (\ref{eq20}) wil allow us to
discuss the current generated by a system of $N_e << N$ electrons.

\subsubsection{Single electron current}\label{Single electron current}

It is instructive to briefly discuss the effect of correlated- and
uncorrelated disorder on the persistent current for a single
electron on the lowest level ($k=0$) of the energy band (\ref{eq8})
for $-1/2<\phi/\phi_0<1/2$.  From (\ref{eq20}) and (\ref{eq6}) we
obtain

\begin{equation}\label{eq22}
\langle I_{k=0}\rangle_{\text{disorder}}= I_0 \Biggl (
\frac{N}{4\pi}\Biggr)^2 \frac{\phi_0}{\phi} \Biggl [
\frac{\varepsilon^2_0}{\sin^2(2\pi\phi/\phi_0)}
+\frac{\varepsilon^2_0\phi_0}{2\pi\phi}\cot\frac{2\pi\phi}{\phi_0}-
2(\varepsilon^2_0+3\alpha^2)\Biggl
(\frac{\phi_0}{2\pi\phi}\Biggr)^2\Biggr ]
 \quad ,
\end{equation}
where $I_0=(4\pi^2 e)/Nh$.

\noindent It follows from (\ref{eq22}) that the correlated
disorder generally leads to an enhancement of persistent current
with respect to the pure system result, $I_{k=0}=-I_0 2\phi/N
\phi_0$, obtained from (\ref{eq8}). In contrast the corresponding
expression for the current induced by uncorrelated disorder
vanishes for $\phi/\phi_0\rightarrow 0$ and leads to suppression
of persistent current at larger flux within the range
(\ref{eq16}).

\subsubsection{Low band filling}\label{low band filling}

At low band filling the occupied energy levels in the ring below
the Fermi level correspond to small $q_k$-values (for large $N$),
so that the terms involving $\sin q_k$ and related small
quantities in denominators dominate in the disorder perturbation
(\ref{eq14}-\ref{eq15}).  By expanding the denominators in
question to leading order in $q_k$ and collecting terms we
obtained the expression (\ref{eq20}) for $\langle
E_k^{(2)}\rangle$ which we now use for calculating the persistent
current induced by the system of $N_e$ electrons in the ground
state when $1\ll N_e\ll N$.  The averaged persistent current from
the $k$-th level is given by

\begin{equation}\label{eq23}
\langle I_{k}\rangle_{\text{disorder}}= \frac{\pi e}{2h}\frac{1}{N
q_k} \Biggl [\frac{\varepsilon^2_0 N}{\sin^2(2\pi\phi/\phi_0)}
+\frac{\varepsilon^2_0}{q_k}\cot\frac{2\pi\phi}{\phi_0}-\frac{2(\varepsilon^2_0+3\alpha^2)}{N
q^2_k}\Biggr ] \quad .
\end{equation}
\noindent A simple analytic procedure for performing the summation
of the currents (\ref{eq23}) from the $N_e$, lowest levels of the
energy band (\ref{eq9}) is to replace the summation by an
integration. For this purpose we use the series summation formula
of Euler-Maclaurin\cite{19},

\begin{equation}\label{eq24}
\sum^{k_+}_{k=k_-}f_k = \int^{K_+}_{K_-}f(k) d\; k-\frac{1}{2}
[f(K_-)+f(K_+)], K_-=k_--1,K_+=k_++1\quad ,
\end{equation}
where we have retained only the leading boundary term of the
general formula\cite{19}.  Here $k_-$ and $k_+$ denote,
respectively, the smallest and the largest value of the quantum
number $k$ labelling the occupied levels in the energy band of the
ring.  From the spectra of occupied energy levels detailed above
for the domains $-1/2<\phi/\phi_0<1/2$ for odd and for even $N_e$
we have: $k_+=-k_-=(N_e-1)/2$ for odd $N_e$; $k_+=N_e/2 -1$ and
$k_-=-N_e/2$ for even $N_e$ with $\phi > 0$; and $k_+=N_e/2,
k_-=-N_e/2\;+1$ for even $N_e$ and $\phi < 0$. We thus obtain the
following values for $q_k$ at the lower and upper limits of the
integral in (\ref{eq24}):

\begin{align}\label{eq25}
q_{K_-} &=\frac{\pi}{N}
\left(-N_e-1+\frac{2\phi}{\phi_0}\right),\quad q_{K_+}
=\frac{\pi}{N}\left(N_e+1+\frac{2\phi}{\phi_0}\right)\quad,\nonumber\\
&(\text{odd
}N_e,-\frac{1}{2}<\frac{\phi}{\phi_0}<\frac{1}{2})\quad ,
\end{align}

\begin{subequations}\label{eq26}
\begin{align}
q_{K_-} &=\frac{\pi}{N}
\left(-N_e+\frac{2\phi}{\phi_0}\right),\quad q_{K_+}
=\frac{\pi}{N}\left(N_e+2+\frac{2\phi}{\phi_0}\right)\quad ,\nonumber\\
&(\text{even }N_e,-\frac{1}{2}<\frac{\phi}{\phi_0}<0)\label{eq26a}\\
 q_{K_-}
&=\frac{\pi}{N} \left(-N_e-2+\frac{2\phi}{\phi_0}\right),\quad
q_{K_+}
=\frac{\pi}{N}\left(N_e+\frac{2\phi}{\phi_0}\right)\quad,\nonumber\\
&(\text{even }N_e,0<\frac{\phi}{\phi_0}<\frac{1}{2})\quad
.\label{eq26b}
\end{align}
\end{subequations}
which are valid for arbitrary $0<N_e<N$.  Having discussed the
case $ N_e=1$ in \ref{Single electron current} above we now focus
mainly on values $N_e\gg 1$ for which

\begin{equation}\label{eq27}
q_{K_+}\simeq -q_{K_-}=\frac{\pi N_e}{N} \Biggl
(1+O\biggl(\frac{1}{N_e}\biggr)\Biggr),\quad N_e\gg 1 \quad ,
\end{equation}
in the various cases detailed in (\ref{eq25}-\ref{eq26}). This
will be sufficient indeed for illustrating the qualitative
differences between these cases.

The total persistent current (\ref{eq7}) obtained from
(\ref{eq23}) involves three types of sums over occupied levels,
namely $S_1=\sum_k q^{-1}_k, S_2=\sum_k q^{-2}_k $ and $
S_3=\sum_k q^{-3}_k $.  We recall the occupied levels in the
various domains of flux, namely $k=-(N_e-1)/2, -(N_e-1)/2\;
+1,\ldots,0,1,\ldots (N_e-1)/2$ for odd $N_e$ and $-1/2<
\phi/\phi_0<1/2, k=-N_e/2\; +1,-N_e/2\; +2,\ldots, 0,\ldots
N_e/2\; -1, N_e/2$ for even $N_e$ and $-1/2<\phi/\phi_0<0$, and
finally, $k=-N_e/2, -N_e/2\; +1,\ldots,0,\ldots N_e/2\; -1$, for
even $N_e$ and $0<\phi<1/2$.  The series $S_1$ which does not
converge for $k\rightarrow\infty$ requires special care.  Using
(\ref{eq12}), we rewrite it in the following forms:

\begin{equation}\label{eq28}
S_1 =\frac{N\phi_0}{2\pi\phi}(1-2J),\quad \text{for odd }
N_e,\quad -\frac{1}{2}<\frac{\phi}{\phi_0}<\frac{1}{2}\quad ,
\end{equation}

\begin{subequations}\label{eq29}
\begin{align}
S_1 &=\frac{N\phi_0}{2\pi\phi}(1-2J)+\frac{1}{q_{k_+}},\quad
\text{for even } N_e,\quad -\frac{1}{2}<\frac{\phi}{\phi_0}<0\quad
,\label{eq29a}\\
S_1 &=\frac{N\phi_0}{2\pi\phi}(1-2J)+\frac{1}{q_{k_-}},\quad
\text{for even } N_e,\quad 0<\frac{\phi}{\phi_0}<\frac{1}{2}\quad
, \label{eq29b}
\end{align}
\end{subequations}

\begin{equation}\label{eq30}
J=\sum^{p}_{k=1} \frac{1}{k^2-(\phi/\phi_0)^2} \quad ,
\end{equation}
with $p =k_+$ in the case of (\ref{eq28}), $p=k_+-1$ in the case
of (\ref{eq29a}) and $p=k_+$ in the case of (\ref{eq29b}).

In applying (\ref{eq24}) for performing the summation over the
occupied levels in (\ref{eq30}) we encounter a difficulty which is
the fact that the integral is not defined since its lower limit
(zero) leads to an imaginary term.  This difficulty may be
resolved by introducing a small $k$-cutoff for the domain of
integration, say at $k\equiv k_c=1$.  This is justified since the
lowest discrete levels contributing to the integral are the levels
$k=\pm 1$, the effect of the level $k=0$ having been separated out
in the first term on the right hand side of
(\ref{eq28},\ref{eq29a},\ref{eq29b}).  With this regularization we
obtain

\begin{equation}\label{eq31}
J=\frac{\phi_0}{2\phi} \Biggl [\ln\Biggl
(\frac{1-\phi/\phi_0}{1+\phi/\phi_0}\Biggr)+\frac{4\phi}{N_e\phi_0}\Biggr]-\frac{\phi^2_0}{2(\phi^2_0-\phi^2)}+O\Bigl
(\frac{1}{N_e^2}\Bigr)\quad .
\end{equation}
On the other hand by summing $S_2$ and $S_3$ using (\ref{eq24}),
we obtain

\begin{equation}\label{eq32}
S_2 =-\frac{N}{2\pi} \Biggl ( \frac{1}{q_{K_+}}-\frac{1}{q_{K_-}}
\Biggr) -\frac{1}{2} \Biggl (
\frac{1}{q^2_{K_+}}+\frac{1}{q^2_{K_-}} \Biggr)\quad ,
\end{equation}

\begin{equation}\label{eq33}
S_3 =-\frac{N}{4\pi} \Biggl (
\frac{1}{q^2_{K_+}}-\frac{1}{q^2_{K_-}} \Biggr) -\frac{1}{2}
\Biggl ( \frac{1}{q^3_{K_+}}+\frac{1}{q^3_{K_-}} \Biggr) \quad ,
\end{equation}
and using the definition (\ref{eq25}) and
(\ref{eq26a},\ref{eq26b}) the leading explicit forms of
(\ref{eq28}), (\ref{eq29a},\ref{eq29b}) and
(\ref{eq32}-\ref{eq33}) are given by:

\begin{equation}\label{eq34}
S_1 =\frac{N\phi_0}{2\pi\phi} (1-2J), \quad\text{odd }N_e,\quad
\end{equation}

\begin{subequations}\label{eq35}
\begin{align}
S_1 &= \frac{N\phi_0}{2\pi\phi} (1-2J) +\frac{N}{\pi N_e} \Biggl[
1+O\Bigl(\frac{1}{N_e}\Bigr) \Biggr] ,\quad \text{even
}N_e,\quad-\frac{1}{2}<\frac{\phi}{\phi_0}<0
 \quad ,\label{eq35a}\\
S_1 &= \frac{N\phi_0}{2\pi\phi} (1-2J) -\frac{N}{\pi N_e} \Biggl[
1+O\Bigl(\frac{1}{N_e}\Bigr) \Biggr] ,\quad \text{even }N_e,\quad
0<\frac{\phi}{\phi_0}<\frac{1}{2} \quad ,\label{eq35b}
\end{align}
\end{subequations}

\begin{equation}\label{eq36}
S_2 =-\frac{N^2}{\pi^2N_e}\Biggl[ 1+O\Bigl(\frac{1}{N_e}\Bigr)
\Biggr], \quad ,
\end{equation}

\begin{equation}\label{eq37}
S_3 =\frac{2\phi}{\phi_0}\Biggl(
 \frac{N}{\pi N_e}
\Biggr)^3 \Biggl[ 1+O\Bigl(\frac{1}{N_e}\Bigr) \Biggr]\quad
,\end{equation} where the limiting forms (\ref{eq36}) and
(\ref{eq37}) are valid for even- as well as for odd $N_e$, in the
domain $-1/2<\phi/\phi_0< 1/2$.

Finally, using (\ref{eq23}) and (\ref{eq34}-\ref{eq37}), the
change in the persistent current (\ref{eq7}) due to the disorder
is given by

\begin{equation}\label{eq38}
\langle I\rangle_{\text{disorder}}= \frac{I_0 N^3
\varepsilon^2_0}{8\pi N_e} \Biggl\{ \Biggl[
\frac{\phi_0}{2\pi\phi} (1-2J)+Q \Biggr] \frac{1}{\sin^2
(2\pi\phi/\phi_0)}  -\frac{1} {\pi^2 N_e }
\cot\frac{2\pi\phi}{\phi_0} + O \Bigl ( \frac{1}{N_e^2}
\Bigr)\Biggr\}\quad ,
\end{equation}
where $Q=0$ for odd $N_e$, $Q=(\pi N_e)^{-1}$ for even $N_e$ with
$-1/2 < \phi/\phi_0<0$ and $ Q=-(\pi N_e)^{-1}$ for even $N_e$
with $0<\phi/\phi_0<1/2$.

It follows from (\ref{eq38}) that for $1\ll N_e\ll N$, the effect
of the disorder is dominated by the form of $J$ in (\ref{eq31}),
which is negative for both signs of the magnetic flux $\phi$. This
shows that for odd $N_e$ the disorder reduces the persistent
current with respect to the pure system result (\ref{eq16}) while
enhancing it with respect to the corresponding pure system values
(\ref{eq17a}) and (\ref{eq17b}) for even $N_e$.

We conclude that while the correlation effect may efficiently offset
the effect of an uncorrelated potential on the persistent current
both in the case of a single electron (see \ref{Single electron
current}) and in that of a gas of a large number of electrons (see
the case $N_e=N/2$ in \ref{half filled band} below), it does not
affect the effect of uncorrelated disorder on the persistent current
in systems with $1<<N_e<<N$ at low order.
\subsection{HALF-FILLED BAND}\label{half filled band}

In the case of a half-filled band the terms proportional to $1/N$
and to $1/N^2$ in the energy level perturbations
(\ref{eq14}-\ref{eq15}) may be ignored relative to the remaining
terms of order one for energy levels (\ref{eq8}) near the Fermi
level.  In this case we thus have from (\ref{eq14}-\ref{eq15})

\begin{equation}\label{eq39}
\langle E^{(2)}_k\rangle^{\text{uncorr}}\simeq
\frac{\varepsilon^2_0}{4\sin q_k}\cot\frac{2\pi\phi}{\phi_0}\quad
,
\end{equation}

\begin{equation}\label{eq40}
\langle E^{(2)}_k\rangle^{\text{corr}}\simeq
\frac{\alpha^2}{2}\Bigl(\cos q_k-\cot\frac{2\pi\phi}{\phi_0}\sin
q_k\Bigr)\quad .
\end{equation}
\noindent The persistent current due to the effect of the
uncorrelated random potential (\ref{eq39}) has been discussed in
Ref. \cite{14} for a half-filled band for odd $N_e$ in the range
$-1/2 < \phi/\phi_0<1/2$ and for even $N_e$ in the range
$0<\phi/\phi_0<1$, using the Euler- Maclaurin series summation
formula. \noindent However, throughout the present work we
consider the typical domain $-1/2 <\phi/\phi_0<1/2$ for the
detailed analyses of the effects of correlated disorder in the
persistent current. This seems more useful since recent first
numerical results for persistent currents in disordered rings with
a correlated random potential have been presented for the range
$-1/2 <\phi/\phi_0<1/2$ only\cite{15,16,17}.  From (\ref{eq39})
the effect of the uncorrelated random potential on the persistent
current in the $k$-th level is given by

\begin{equation}\label{eq41}
\langle I_k\rangle_{\text{uncorr}}= \frac{\pi
e}{2h}\frac{\varepsilon^2_0}{\sin^2 (2\pi\phi/\phi_0)}
\frac{1}{\sin q_k}+O (1/N)\quad .
\end{equation}
whose summation for the $N_e$ lowest occupied levels of the energy
band (\ref{eq8}) in the domains $-1/2 <\phi/\phi_0<0$ and $0
<\phi/\phi_0<1/2$, respectively is performed using the Euler-
Maclaurin formula (\ref{eq24}).  Using (\ref{eq18}), this yields
the following final result for the effect of the uncorrelated
disorder on the mean persistent current in the half-filled band:

\begin{multline}\label{eq42}
\langle I\rangle_{\text{uncorr}}
=I_0\frac{N\varepsilon^2_0}{8\sin^2(2\pi\phi/\phi_0)} \Biggl [
-\frac{N}{4\pi}\ln \frac {(1+\cos q_{K_+})(1-\cos q_{K_-})}
{(1-\cos q_{K_+})(1+\cos q_{K_-})}\Biggr .
\\
\Biggl . -\frac{1}{2} \Biggl ( \frac{1}{\sin q_{K_-}}+
\frac{1}{\sin q_{K_+}} \Biggr) \Biggr]\quad ,
\end{multline}
in terms of the $q_k$- values (\ref{eq25},\ref{eq26a},\ref{eq26b})
for $N_e=N/2$, near the upper and lower boundary values $k_+$ and
$k_-$ of the domain of occupied $k$-levels and for flux values
within the interval ($-\phi_0/2, \phi_0/2$).  The equation
(\ref{eq42}) may be readily simplified for $N_e=N/2\gg 1$ by
expanding the square bracket to leading order in $1/N$, using
(\ref{eq25},\ref{eq26a},\ref{eq26b}).  This yields

\begin{equation}\label{eq43}
\langle I\rangle_{\text{uncorr}}
=I_0\frac{N\varepsilon^2_0}{8\sin^2(2\pi\phi/\phi_0)} A\quad ,
\end{equation}

\begin{equation}\label{eq44}
A= \frac{2\phi}{\phi_0}+O(1/N^2),\quad\text{odd }N_e,\quad
-\frac{1}{2}<\frac{\phi}{\phi_0}<\frac{1}{2}\quad,
\end{equation}

\begin{subequations}\label{eq45}
\begin{align}
A &= 1+ \frac{2\phi}{\phi_0}+ O(1/N^2),\quad\text{even }N_e,\quad
-\frac{1}{2}<\frac{\phi}{\phi_0}<0,\quad ,\label{eq45a}\\
A &= -1+ \frac{2\phi}{\phi_0} + O(1/N^2),\quad\text{even
}N_e,\quad 0<\frac{\phi}{\phi_0}<\frac{1}{2}, \quad ,\label{eq45b}
\end{align}
\end{subequations}
where (\ref{eq43}) with $A$ defined by (\ref{eq44}) or
(\ref{eq45b}) has been obtained earlier in Ref. \cite{14}.

The persistent current due to the correlation for an electron on
the $k$-th level is, from (\ref{eq6}) and (\ref{eq40}),

\begin{equation}\label{eq46}
\langle I_k\rangle_{\text{corr}} =-\frac{\pi
e}{h}\frac{\alpha^2}{\sin^2 (2\pi\phi/\phi_0)} \sin q_k +
O(1/N)\quad ,
\end{equation}
which leads to the following exact results for the persistent
current for the system of $N_e$ spinless electrons occupying the
lower half of the energy band of the system:

\begin{align}\label{eq47}
\langle I\rangle_{\text{corr}} &= -I_0 \frac{N
\alpha^2}{4\sin^2(2\pi\phi/\phi_0)}
\frac{\sin(2\pi\phi/N\phi_0)}{\sin\pi/N}
\nonumber\\
& \simeq -I_0 \left( \frac{2\phi}{\phi_0}\right)
\frac{ N \alpha^2}{4\sin^2(2\pi\phi/\phi_0)}\quad ,\nonumber\\
& (\text{odd }N_e, -1/2<\phi/\phi_0 <1/2)\quad ,
\end{align}

\begin{align}\label{eq48}
\langle I\rangle_{\text{corr}} &= -I_0 \frac{N
\alpha^2}{4\sin^2(2\pi\phi/\phi_0)}
\frac{\sin\frac{\pi}{N}\left(\frac{2\phi}{\phi_0}\pm
1\right)}{\sin\pi/N}
\nonumber\\
& \simeq -I_0 \left( \frac{2\phi}{\phi_0}\pm 1\right)
\frac{ N \alpha^2}{4\sin^2(2\pi\phi/\phi_0)}\quad ,\nonumber\\
& (\text{even }N_e, +\text{sign for }-1/2<\phi/\phi_0 <0,
-\text{sign for }0<\phi/\phi_0 <1/2)\quad ,
\end{align}
\noindent By comparing these expressions with the pure system
currents (\ref{eq16}) and (\ref{eq17a},\ref{eq17b}) it follows
that the effect of the correlation of the random potential is, in
all cases, to enhance the persistent current.

Finally the total average persistent current $\langle I\rangle$ in
the ring is given by the sum of free electron currents [large
$N$-limits of pure tight-binding system currents, Eqs.
(\ref{eq16})(\ref{eq17a})(\ref{eq17b})] and the changes in the
persistent current induced by the disorder via the correlations of
the random potential [Eqs. (\ref{eq47}-\ref{eq48})] and via the
random potential in the absence of correlation [Eqs.
(\ref{eq43}-\ref{eq45})]:

\begin{equation}\label{eq49}
\langle I\rangle= -I_0 \left ( \frac{2\phi}{\phi_0} \right )
    \left[1-\frac{N(\varepsilon^2_0-2\alpha^2)}{8\sin^2 (2\pi\phi/\phi_0)}\right], \text{odd }N_e,
-\frac{1}{2}<\frac{\phi}{\phi_0}<\frac{1}{2}\quad ,
\end{equation}

\begin{subequations}\label{eq50}
\begin{align}
\langle I\rangle &= -I_0 \left ( \frac{2\phi}{\phi_0}+1 \right )
\left(1-\frac{N(\varepsilon^2_0-2\alpha^2)}{8\sin^2(2\pi\phi/\phi_0)}\right
),
\text{even }N_e, -\frac{1}{2}<\frac{\phi}{\phi_0}<0\quad ,\label{eq50a}\\
\langle I\rangle &= -I_0 \left ( \frac{2\phi}{\phi_0}-1 \right )
\left(1-\frac{N(\varepsilon^2_0-2\alpha^2)}{8\sin^2(2\pi\phi/\phi_0)}\right
), \text{even }N_e, 0<\frac{\phi}{\phi_0}<\frac{1}{2}\quad ,
\label{eq50b}
\end{align}
\end{subequations}
with $I_0$ defined in (\ref{eq18}).  These expressions are invalid
near $\phi=0$, which is a consequence of the breakdown of the
perturbation expression \eqref{eq11} for flux values equal to
half-integer multiples of the flux quantum, as mentioned earlier.

The expressions \eqref{eq49} and (\ref{eq50a}-\ref{eq50b}) indicate
that the disorder acts to renormalize the free particle persistent
currents \eqref{eq16} and (\ref{eq17a}-\ref{eq17b} by a common
factor.  In particular, it follows that the correlation of the
random potential leads to a systematic increase of the persistent
current proportional to $\alpha^2$.  However, the important feature
of (\ref{eq50a}-\ref{eq50b}) is that for $\alpha^2$ less than the
critical value $\alpha^2\equiv\alpha^2_c=\varepsilon^2_0/2$ the
persistent current is reduced by the correlated disorder while for
$\alpha^2>\alpha^2_c$ it is enhanced by it for any parity.  In this
respect, the results for the persistent current at low band filling,
$1\ll N_e\ll N$, discussed in \ref{low filling of energy band} above
show an interesting asymmetry.  Indeed, in this case the effect of
the correlated disorder is to reduce the persistent current for odd
$N_e$ and to enhance it for even $N_e$.  This is a consequence of
the fact that the flux dependence of the persistent current
(\ref{eq38}) induced by the disorder does not follow the simple
linear scalings of the pure system currents (\ref{eq16},
\ref{eq17a}, \ref{eq17b}).

\subsection{DISCUSSION OF THE RESULTS}\label{discussion of the results}

In this subsection we discuss the restrictions on ring lengths $N$
and on relative flux values in \eqref{eq49} a nd
(\ref{eq50a}-\ref{eq50b}) which are imposed by our non-degenerate
perturbation study of the effect of the disorder on the energy
levels of the ring in the presence of the applied flux.  We focus
mainly on the case of the half-filled energy band where the effects
of the disorder on the persistent current may be observed in wide
range of nanoscale ring parameters, as shown below.

First, the perturbation expansion \eqref{eq9} in terms of the random
site energies relative to the fixed intersite hopping parameter
implies, by definition,

\begin{equation}\label{eq51}
|\varepsilon_0|<<1, |\alpha|<<1 \quad ,
\end{equation}
which reflects the fact that the first order correction to the
energy levels, $E^{(0)}_k$, of the pure system is small.  Next, the
typical value of the first order shift of energy levels due to the
disorder, $\sqrt{\langle E^{(1)2}_k\rangle}=
(\varepsilon^2_0+\alpha^2)/N$ must be small compared to the spacing,
$\vert \Delta\vert$, of the energy levels of the pure system.  From
\eqref{eq8} we obtain

$$\Delta= E_{k+1}^{(0)}(\phi)-
E_{k}^{(0)}(\phi)= \frac{4\pi}{N}\sin\frac{2\pi}{N}
(k+\frac{1}{2}+\frac{\phi}{\phi_0})\quad ,$$ which leads to the
condition

\begin{equation}\label{eq52}
\sqrt{N} \sqrt{\varepsilon_0^2+\alpha^2}<< 4\pi \sin \frac{2\pi}{N}
\left(\frac{1}{2}+\frac{\phi}{\phi_0}\right) \simeq 4\pi \quad ,
\end{equation}
at the Fermi level (where $k=\pm N_e/2$) in the half-filled band
case.  This condition which is important for close-spaced level
systems (large $N$) expresses that disorder acts as a small
perturbation on the (observable) excitation frequencies defined by
$\Delta$.  Since, for band fillings $N_e<<N$, the Eq. \eqref{eq52}
restricts the discussion to rings of perimeters less than a few
nanometers, we focus, in the remainder of this subsection, on the
half-filled band case $N_e=N/2$, where our analysis covers a much
wider range of nanometric rings sizes.  Also since here we are
mainly interested in analyzing the effect of small random potential
correlations ($\alpha^2\leq\varepsilon^2_0$) for rings of fixed
length $N$ we replace the condition \eqref{eq52} by the slightly
weaker form:

$$\sqrt{N} \sqrt{\varepsilon_0^2}<< 4\pi
\quad\quad\quad\quad\quad\quad\quad\quad\text{(52.a)}$$ at the fermi
level of the half-filled band.

Finally, an important restriction on the allowed flux values in
\eqref{eq49} and \eqref{eq50a}, \eqref{eq50b} arises from the second
order perturbation expression for the effect of the disorder in the
energy levels (\ref{eq9}-\ref{eq11}).  This requires the typical
second order corrections of the energy levels to have magnitudes of
the order of $\varepsilon_0^2$.  In the half-filled band case this
condition is generally obeyed by Eqs (\ref{eq39}-\ref{eq40}) for
$\langle E^{(2)}_k\rangle$ at the fermi level (where $\sin q_k
\simeq 1$) by requiring $\cot \frac{2\pi\phi}{\phi_0}\simeq 1$. This
yields a lower limit

\begin{equation}\label{eq53}
\phi_{\text{min}}=0,125 \phi_0 \quad ,
\end{equation}
for the flux values at which the perturbation treatment of the
disorder is valid.

We close this subsection with some typical numerical results for the
renormalization factor in \eqref{eq49}, \ref{eq50a} and
\eqref{eq50b},

\begin{equation}\label{eq54}
A=
1-\frac{N\varepsilon_0^2(1-2\eta)}{8\sin^2\frac{2\pi\phi}{\phi_0}},
\eta=\frac{\alpha^2}{\varepsilon_0^2} \quad ,
\end{equation}
of the pure system current by the effect of the disorder as a
function of the correlation parameter
$\eta=\alpha^2/\varepsilon_0^2$.  We choose $|\varepsilon_0|=0.1$
and define a typical ring length $N=(0,4\pi)^2/\varepsilon_0^2=158$
compatible with (52.a).  Using these values, we obtain the results
listed in table 1, for various values of $\eta$, for
$\phi=\phi_{\text{min}}$ and for an intermediate value,
$\phi=0.25\phi_0$, in the domain $0<\phi/\phi_0<1/2$.  Note that
increasing the flux in the case $\eta<0.5$ enhances the magnitude of
the disorder effect on the persistent current, while in the case
$\eta>0.5$, increasing the flux reduces the effect of the correlated
disorder.

\section{CONCLUDING REMARKS}\label{concluding remarks}

The main conclusions of this work have been discussed in
Sect.\ref{the persistent current in the ring} along with our
detailed results for persistent currents in a metal ring in the
presence of correlated disorder and the discussion of their domain
of validity.

We conclude with some further general remarks, particularly in
connection with the yet unexplained experimental observations of
persistent currents.

By studying the persistent current in sufficiently large ensembles
of rings it has been possible recently to obtain the correct sign,
in addition to the magnitude, of the current\cite{26,27,21}.  These
measurements have shown the persistent current to be diamagnetic at
low fields (i.e. $\frac{d\langle I\rangle}{d\phi}<0, \phi\rightarrow
0$). This observation constitutes a severe additional constraint for
the validity of theoretical explanations of persistent current in
disordered rings.  Indeed, as shown by \eqref{eq16}, \eqref{eq17a}
and \eqref{eq17b}, a diamagnetic persistent current for any parity
of the number of conduction electrons is generally obtained only for
non-interacting free electrons.  Now, our analytical expressions
\eqref{eq49}, \eqref{eq50a} and \eqref{eq50b} for the effect of
disorder on the persistent current in a half-filled band includes an
additional flux-dependence which prevents the current from having a
fixed negative sign independent of parity and ring size.  We
observe, however, that our theoretical results for persistent
currents are inapplicable to measurements in rings of mesoscale
perimeters of 2 to 8 $\mu$m used in the various experimental
studies\cite{2,3,4} and \cite{26,27,21}.  Indeed, for values
$|\varepsilon_0^2|\simeq 0.1$, whith corresponding values
$0<|\alpha|\leq 0.1$, the condition \eqref{eq52} restricts the
validity of our results to rings with a few hundred atomic sites
i.e. to nanoscale perimeters less than 100 nm.  We note incidently
that typical ring sizes for which the effect of correlations of the
random potential has been analyzed numerically in
Refs.\cite{15,16,17} lie also in this range.  In conclusion, it
would be interesting if experimental studies could be performed for
ensembles of rings of nanoscale sizes in order to test the detailed
behavior of persistent currents shown theoretically in this paper
and numerically, for a different model, in Refs.\cite{15,16,17}.

Our results for the persistent current for a half-filled band in our
ARDM model indicate a strong enhancement of the current due to the
correlation of the random potential for $\alpha^2\geq
\varepsilon_0^2/2$. The detailed mechanism for the enhancement of
persistent current by correlation for the ARDM model\cite{9} on a
ring threaded by a magnetic flux is quite different from the
mechanism identified in recent numerical calculations for the RDM
model of Dunlap, Wu and Phillips\cite{10} in Refs. \cite{15,17}.  We
recall that the linear chain RDM model differs essentially from the
ARDM by the existence of a fraction $\sqrt N$ of extended
states\cite{10,11}. For the RDM model values of persistent current
of the order of the values for free electrons are found only when
the Fermi level coincides with the energy of an extended state, both
for even $N_e$\cite{15,17} and for odd $N_e$ (for an asymmetric
dimer model)\cite{16}.  In contrast, for $N_e$ values corresponding
to relatively large band fillings but for which the Fermi level does
not coincide with an extended state, the persistent current is found
to be strongly diminished by the correlated disorder\cite{15,16}.
Thus for the RDM a strong compensation of the current reduction due
to the uncorrelated disorder by the effect of correlation does not
seem to exist\cite{15,16}, in contrast to what is shown for the
half-filled band in (\ref{eq49}) and (\ref{eq50a},\ref{eq50b}) for
the ARDM.  We note, however, that the numerical results of
Refs\cite{15,16} pertain to a case of maximal disorder (equal number
of $a$- and $b$-atoms in the RDM), while our analytical results are
restricted to weak disorder in the ARDM.

As shown in Sect. \ref{discussion of the results}, our results for
the persistent current are valid only for flux values larger than a
value of the order of $\phi_0/8$.  Thus the sign of the persistent
current obtained form Eqs. (\ref{eq49}, \ref{eq50a}, \ref{eq50b})
could be compared with experimental results at flux values
$\phi>\phi_0/8$ only, rather than for $\phi\rightarrow 0$ as in
recent experimental studies\cite{26,27,21}.

\section{APPENDIX}\label{appendix}
\setcounter{equation}{0}
\renewcommand{\theequation}{A.\arabic{equation}}

Defining
\begin{equation}
\label{eqA1} A_{n\; m}=\frac{\sin(n-m)q_k\sin (N-n+m)q_k}{\sin
q_k\sin N\; q_k} \quad ,
\end{equation}
the various contributions to the disorder average, $\langle
E^{(2)}_k\rangle$, of (\ref{eq11}) are given by

\begin{equation}\label{eqA2}
\frac{1}{N} \sum^N_{n=2}\sum^{n-1}_{m=1} \langle(E^{(1)})^2\rangle
A_{nm} = -\frac{(\varepsilon^2_0+\alpha^2)}{4 N\sin q_k} (N\cot N
q_k -\cot q_k)\quad ,
\end{equation}

\begin{equation}\label{eqA3}
\frac{1}{N} \sum^N_{n=2}\sum^{n-1}_{m=1}
\langle\varepsilon_m\varepsilon_n\rangle A_{nm} = \frac{\alpha^2}{2}
\frac{\sin (N-1)q_k}{\sin q_k} \quad ,
\end{equation}

\begin{multline}\label{eqA4}
-\frac{1}{N} \sum^N_{n=2} \sum^{n-1}_{m=1} \langle\varepsilon_m
E^{(1)}\rangle A_{nm}= \frac{(\varepsilon_0^2-\alpha^2)}{4N\sin q_k}
(N\cot N q_k-\cot q_k)+ \frac{\alpha^2}{4N^2\sin q_k\sin N q_k}
\Biggl\{ N^2\cos N q_k
\Biggr.\\
\Biggl. -2\Biggl [ \frac{e^{-i(N+2)q_k}}{1-e^{-4iq_k}} \Biggl (
\frac{e^{4iq_k}(1-e^{2iN q_k})}{1-e^{4iq_k}} -\frac{N}{2} \Biggr )
+ \text{c.c.}
\Biggr ]\\
\Biggl. -2\Biggl [ \frac{e^{-iNq_k}}{1-e^{-4iq_k}} \Biggl (
\frac{1-e^{2iN q_k}}{1-e^{4iq_k}} -\frac{N}{2} \Biggr ) +
\text{c.c.} \Biggr ] \Biggr\}\quad ,
\end{multline}

\begin{multline}\label{eqA5}
-\frac{1}{N} \sum^N_{n=2} \sum^{n-1}_{m=1} \langle
E^{(1)}\varepsilon_n \rangle A_{nm}=
\frac{(\varepsilon_0^2-\alpha^2)}{4N\sin q_k} (N\cot N q_k-\cot
q_k)+ \frac {\alpha^2} {4N^2\sin q_k\sin N q_k} \Biggl\{ 2N
(N-1)\cos N q_k
\Biggr .  \\
-\Biggl [ \frac {e^{-i(N-2)q_k}} {1-e^{-2i q_k}} \Biggl ( \frac
{1-e^{2i(N-1)q_k}} {1-e^{2iq_k}} +\frac {1+e^{-2i q_k}-e^{2i N
q_k}(e^{4i q_k}+e^{-2i q_k})} {1-e^{4i q_k}} \Biggr.
\Biggr. \\
-(N-1)e^{2iq_k}- \biggl(\frac{N}{2}+1\biggr) e^{-2iq_k}-
\frac{N}{2} e^{-6i q_k} \Biggr ) +\text{c.c.} \Biggr]
\Biggr\}\quad .
\end{multline}

The sum of (\ref{eqA2}-\ref{eqA5}) leads after appropriate
reductions to the final form in (\ref{eq13}-\ref{eq15}).

\begin{table}[htbp]
\begin{center}
\begin{tabular}{|c|c|c|}
\hline
$\eta$ & $\frac{\phi}{\phi_0}$ & A \\
\hline 0 & 0,125 & 0,605\\
 & 0,25 & 0,803\\
 \hline
 0,2 & 0,125 & 0,769\\
 & 0,25 & 0,882\\
  \hline
 0,4 & 0.125 &0,921 \\
 & 0.25 & 0,961\\
  \hline
    0,5 & - & 1\\
    \hline
  0,6 & 0,125 & 1,079\\
  & 0,25 & 1,040\\
  \hline
\end{tabular}
\end{center}
\caption{Disorder dependent free electron persistent current
renormalization factor \eqref{eq53} for various values of the
disorder correlation parameter $\eta$}
\end{table}


\begin{thebibliography}{unsrt}
\bibitem{23} F. London, J. Phys. Radium {\bf 8}, 397 (1937).
\bibitem{24} F. Hund, Ann. Phys. (Leipzig), {\bf 32}, 102 (1938).
\bibitem{1} M. B\"{u}ttiker, Y. Imry and R. Landauer, Phys. Lett. {\bf 96A}, 395 (1983).
\bibitem{2} L.P. L\'{e}vy, G. Dolan, J. Dunsmuir and H. Bouchiat, Phys. Rev. Lett. {\bf 64}, 2074 (1990).
\bibitem{3} V. Chandrasekhar, R.A. Webb, M.J. Brady, M.B. Ketchen, W.J. Galager and A. Kleinsasser, Phys. Rev. Lett. {\bf 67}, 3578 (1991).
\bibitem{4} D. Mailly, C. Chapelier and A. Benoit, Phys. Rev. Lett. {\bf 70}, 2020 (1993).
\bibitem{5} Y. Imry, Introduction to Mesoscopic Physics (Oxford University Press, Oxford, 1997).
\bibitem{6} K. Efetov, Supersymmetry in Disorder and Chaos (Cambridge University Press, Cambridge, 1997).
\bibitem{7} U. Eckern and P. Schwab, Adv. Phys. {\bf 44}, 387
(1995); ibid. J. Low Temp. Phys. {\bf 126}, 1291 (2002).
\bibitem{8} M. Abraham and R. Berkovits, Phys. Rev. Lett. {\bf 70}, 1509 (1993); G. Bouzevar, D. Poilblanc and G. Montambaux, Phys. Rev. B{\bf 49},8258 (1994); H. Kato and D. Yoshioka, Phys. Rev. B{\bf 50}, R4943 (1994), J.-X. Zhu, Z.D. Wang and L. Sheng, Phys. Rev. B{\bf 52}, 14505 (1995).
\bibitem{9} J. Heinrichs, Phys. Rev. B{\bf 51}, 5699 (1995); Phys. Rev. B {\bf 52}, 15649 (E) (1995).
\bibitem{10} D.H. Dunlap, K. Kundu and P. Phillips, Phys. Rev. B{\bf 40}, 10999 (1989); D.H. Dunlap, H.L. Wu and P. Phillips, Phys. Rev. Lett. {\bf 65}, 88 (1990);
\bibitem{11} P. Phillips, Advanced Solid State Physics (Westview Press, Oxford, 2003).
\bibitem{12} M. Kappus and F. Wegner, Z. Phys. B{\bf 45}, 15 (1981).
\bibitem{13} H.F. Cheung, Y. Gefen, E.K. Riedel and W.H. Shih, Phys. Rev. B{\bf 37}, 6050
(1988); ibid. IBM J. Res. Develop. {\bf 32}, 359 (1988).
\bibitem{14} J. Heinrichs, Int. J. Mod. Phys. B{\bf 16}, 593 (2002).
\bibitem{15} Y.M. Liu, R.W. Peng, X.Q. Huang, Mu Wang, A. Hu and S.S. Jiang, J. of the Phys. Soc. of Japan, {\bf 72}, 346 (2003).
\bibitem{16} X.F. Hu, Z.H. Peng, R.W. Peng, Y.M. Liu, F. Liu, X.Q. Huang, A. Hu and S.S. Jiang, J. of Applied Physics, {\bf 95}, 7545 (2004).
\bibitem{17} X. Chen, Z.Y. Deng, W. Lu and S.C. Shen, Phys. Rev. B{\bf 61}, 2008 (2000).
\bibitem{18} I. Tomita, Physica A {\bf 308}, 1 (2002).
\bibitem{22} General expressions for energy levels perturbed by a
weak disorder similar to (8-12) have been derived earlier in the
case of a ring threaded by a non-hermitian field in J. Heinrichs,
Phys. Rev. B{\bf 63}, 165108 (2001).  The interest in this probelm
of non-hermitian quantum mechanics arose from the study of the
pinning of vortices by columnar defects in a superconductor.
\bibitem{25} A. Altland, S. Ida, A. M\"{u}ller-Groeling and H.A.
Weidenm\"{u}ller, Europhys. Lett. {\bf 20}, 155 (1992).
\bibitem{19} M. Abramovitz and I.A.Stegun, Handbook of Mathematical Functions, Chap. 3 (National Bureau of Standards, Washington, 1965).
\bibitem{26} P. Mohanty, Ann. Phys. (Leipzig), {\bf 8}, 549 (1999).
\bibitem{27} E.M.Q. Jariwala, P. Mohanty, M.B. Ketchen and R.A.
Webb, Phys. Rev. Lett. {\bf 86}, 1594 (2001).
\bibitem{21} R. Deblock, R. Bel, B. Beulet, H. Bouchiat, and D. Mailly, Phys. Rev. Lett. {\bf 89}, 206803 (2002).
\end{thebibliography}
\end{document}